\documentclass[superscriptaddress,prd,twocolumn]{revtex4}
\usepackage{latexsym}
\usepackage{amsthm}
\usepackage{amsmath}
\usepackage{graphicx}
\usepackage{amssymb}
\usepackage{bbm}
\usepackage{slashed}
\newcommand{\gsim}{\lower.7ex\hbox{$\;\stackrel{\textstyle>}{\sim}\;$}}
\newcommand{\lsim}{\lower.7ex\hbox{$\;\stackrel{\textstyle<}{\sim}\;$}}

\def\epm{e^+e^-}
\newcommand{\cm}{~\mathrm{cm}}
\newcommand{\exE}{\delta}

\newcommand{\mX}{m_{\scriptscriptstyle \chi}}

\newcommand{\eqtau}{\tau_{\scriptscriptstyle {\rm eq}}}

\newcommand{\TeV}{\,\mathrm{TeV}}
\newcommand{\GeV}{\,\mathrm{GeV}}
\newcommand{\MeV}{\,\mathrm{MeV}}
\newcommand{\keV}{\,\mathrm{keV}}

\newcommand{\ab}{\,\mathrm{ab}}

\newcommand{\pb}{\,\mathrm{pb}}

\newcommand{\Br}{{\text{ Br}}}

\newcommand{\be}{\begin{equation}}
\newcommand{\ee}{\end{equation}}
\newcommand{\bea}{\begin{eqnarray}}
\newcommand{\eea}{\end{eqnarray}}
\newcommand{\f}[2]{\ensuremath{\frac{#1}{#2}}}

\newcommand{\bef}{\begin{figure}[htbp]\begin{center}}
\newcommand{\eef}{\end{center}\end{figure}}

\begin{document}
\pagestyle{plain}
\title{
\begin{flushright}
\mbox{\normalsize SLAC-PUB-13807}\\
\mbox{\normalsize SU-ITP-09/43} \\
\end{flushright}
Terrestrial and Solar Limits on Long-Lived Particles in a Dark Sector}
\author{Philip Schuster}
\affiliation{Theory Group, SLAC National Accelerator Laboratory, Menlo Park, CA 94025}
\author{Natalia Toro}
\affiliation{Theory Group, Stanford University, Stanford, CA 94305}
\author{Itay Yavin}
\affiliation{Center for Cosmology and Particle Physics, New York University, New York, NY 10003}
\date{\today}
\begin{abstract}
Dark matter charged under a new gauge sector, as motivated by recent
data, suggests a rich GeV-scale ``dark sector'' weakly coupled to the
Standard Model by gauge kinetic mixing.  The new gauge bosons can
decay to Standard Model leptons, but this mode is suppressed if decays
into lighter ``dark sector'' particles are kinematically allowed.
These particles in turn typically have macroscopic decay lifetimes
that are constrained by two classes of experiments, which we discuss.
Lifetimes of $10 \cm \lesssim c\tau \lesssim 10^{8} \cm$ are
constrained by existing terrestrial beam-dump experiments.  If, in
addition, dark matter captured in the Sun (or Earth) annihilates into
these particles, lifetimes up to $\sim 10^{15} \cm$ are constrained by
solar observations.  These bounds span fourteen orders of magnitude in
lifetime, but they are not exhaustive.  Accordingly, we identify
promising new directions for experiments including searches for
displaced di-muons in B-factories, studies at high-energy and
-intensity proton beam dumps, precision gamma-ray and electronic
measurements of the Sun, and milli-charge searches re-analyzed in this
new context.
\end{abstract}
\maketitle

\section{Light Long-Lived Particles from New Gauge Sectors}

Recent astrophysical anomalies have motivated models
\cite{Finkbeiner:2007kk,Pospelov:2007mp,Cholis:2008vb,ArkaniHamed:2008qn,Pospelov:2008jd}
of TeV-scale dark matter interacting with new states at the GeV scale.
A compelling candidate is a new gauge force, mediated by a $\GeV$-mass
vector-boson that mixes kinetically with hypercharge:
\bea
\label{eq:Lag} 
\mathcal{L} &=&  \mathcal{L}_{\textrm{{SM}}}  +
i\bar{\chi}\left(\slashed{\partial} + i g \slashed{A'}\right) \chi +
M_\chi\bar{\chi}\chi \\\nonumber 
&-&\frac{1}{4}F'^{,\mu\nu}F'_{\mu\nu}+ \frac{\epsilon}{2 \cos\theta_W}
F'_{\mu\nu} B^{\mu\nu} + \mathcal{L}_{\textrm{{Dark}}}.
\eea
A program of searches in high-luminosity $\epm$ collider data
\cite{Essig:2009nc,Batell:2009yf,Reece:2009un}, new fixed-target
experiments \cite{Bjorken:2009mm,Batell:2009di}, and high-energy
colliders \cite{ArkaniHamed:2008qp,Baumgart:2009tn} can search quite
exhaustively for the new gauge boson $A'$ if it decays into a lepton
pair through the $\epsilon$-suppressed kinetic mixing term.

The $A'$ can, however, be accompanied by ``dark sector'' particles
(other vectors, Higgs-like scalars, pseudoscalars, or fermions)
to which it couples without $\epsilon$-suppression, so that the decays of $A'$ into dark sector particles dominate. These particles may decay
to Standard Model matter (and indeed some \emph{must}, if dark matter
annihilation is to explain the astrophysical anomalies),
but unlike the $A'$ their lifetimes are typically macroscopic ---
we will call such particles ``long-lived particles'' (LLP's).

\begin{figure}
\includegraphics[width=0.45\textwidth]{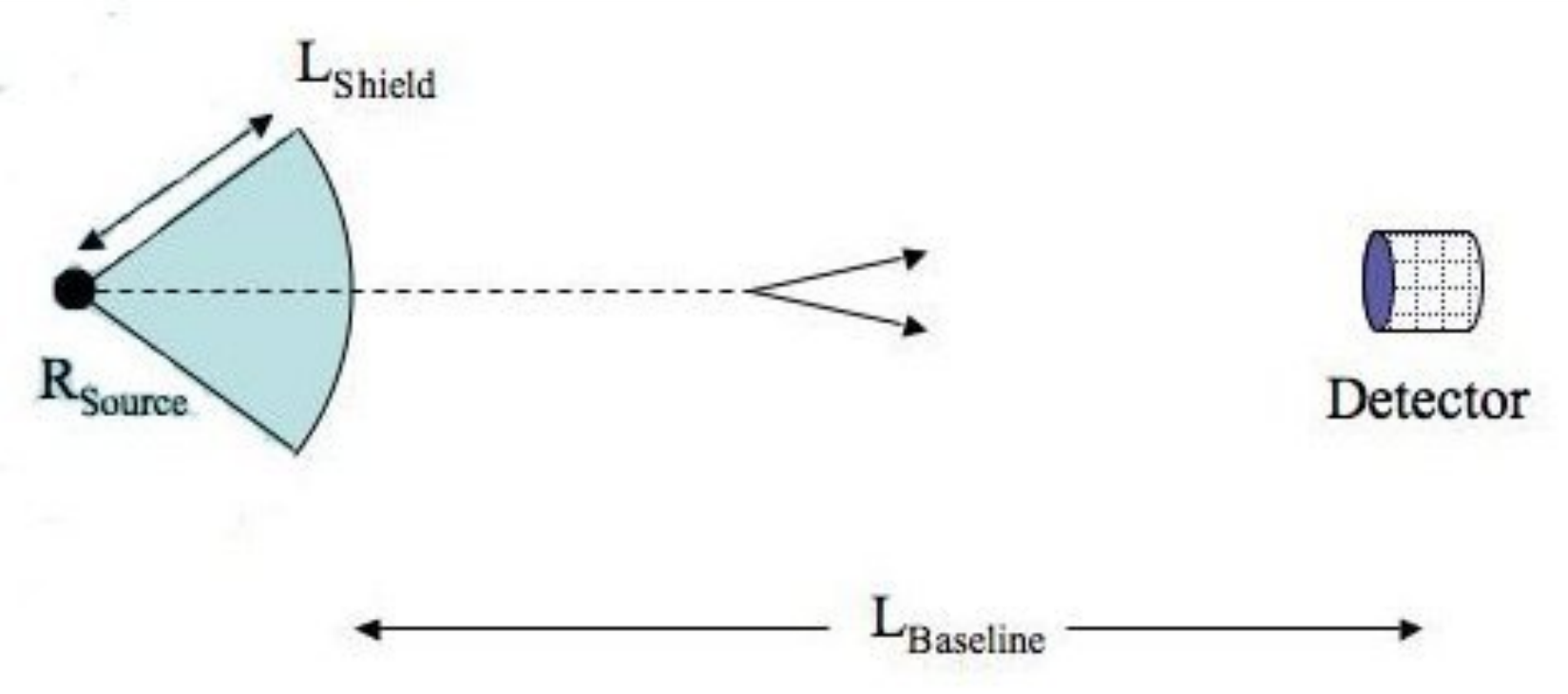}
\caption{A pictorial summary of the two similar approaches discussed in this paper: A strong source, either a beam dump or dark matter annihilation in the Sun, produces dark sector particles. LLP's can penetrate the shield, either the Sun's interior or the beam dump shielding. Decays downstream over a baseline, $L_{\mbox{Baseline}}$, can be detected.}
\label{fig:Setup}
\end{figure}

In this paper, we discuss observational constraints on the decays of the LLPs which cover lifetimes spanning fourteen orders of magnitude, $10 \cm \lesssim c\tau \lesssim 10^{15} \cm$. A schematic diagram illustrating the type of experiments we consider is shown in Fig. \ref{fig:Setup}. The constraints fall into two categories:
\begin{description}
\item[Terrestrial beam dump experiments] (Sec.~\ref{sec:beamDump}): 
 Experiments at proton and electron beam dumps search for decays of long-lived particles in an instrumented region downstream of a thick shield.  Though $A'$ production scales as  $\epsilon^2$, the integrated luminosities $\sim10^2\ab^{-1}$ allow searches for LLP's with mass $\lsim\rm{few }\GeV$ and $10 \lesssim c\tau \lesssim 10^8$ cm.
\item[Annihilation of Dark Matter in the Sun] (Sec. ~\ref{sec:solar}):
  Dark matter captured in the Sun through $A'$-mediated scattering
  annihilates efficiently, with a rate scaling as $\epsilon^2$. Annihilations into LLP's lighter than the $\tau$ threshold result in
  $e$, $\mu$, and light hadrons, which can be detected when the
  LLP decays outside the Sun but before passing the Earth.  In this case, neutrino,
  electron, and gamma ray observations together constrain dark matter
  annihilation into LLP's with $10^7 \cm \lesssim c\tau \lesssim
  10^{15}$ cm.
\end{description}
In Sections \ref{sec:beamDump} and \ref{sec:solar}, we describe the
best available limits.  In Section \ref{sec:Example}, we interpret
these limits in a particular illustrative case: a Higgs-like scalar in
the dark sector that decays into two Standard Model fermions.  We
conclude in Section \ref{sec:discussion}, and highlight the value of
direct $A'\rightarrow \ell^+\ell^-$ searches, searches for displaced
di-muons in B-factories, high-energy and high-intensity proton beam dumps,
precision gamma-ray and electronic measurements of the Sun, and
milli-charge searches in the context of these models.

 \subsection{Decay Lengths in Dark Sectors}\label{sec:orientation} 
Before proceeding, we summarize the parametric scaling that governs the decays of various particles in a dark
sector which is coupled to the Standard Model through gauge kinetic mixing.

In non-Abelian dark sectors, multiple gauge bosons can mix with one
another \cite{ArkaniHamed:2008qn,Baumgart:2009tn}, so that each
acquires an effective kinetic mixing $\epsilon_{eff}$ and lifetime
$\propto \epsilon_{eff}^{-2}$.  Particles $\chi$ of other spins can have
three-body decays into Standard Model charged matter ($\chi \rightarrow f^+ f^- \chi'$ through an off-shell $A'$), with lifetimes $\propto \epsilon^{-2}$ but longer than those of vectors due to
three-body phase space.  For $\epsilon \sim 10^{-2}-\mbox{ few}\times10^{-5}$ and
moderate mass hierarchies these lifetimes are less than 1 m, and multi-body final states can be observed in a collider detector.
B-factory searches in higher-multiplicity final states (e.g. \cite{Aubert:2009pw}) can be
quite sensitive to these decay modes \cite{Batell:2009yf,Essig:2009nc}.

Much longer lifetimes can arise from $\epsilon^{-4}$ scaling or decays
through higher-dimension operators --- these are the modes of greatest
interest for our discussion.  In particular, we consider a Higgs-like
boson $h_D$ below the $A'$ mass.  The $A'$ can decay into $h_D$ only
in models with multiple dark-sector Higgses, as the second decay
product must be a pseudoscalar.  Several different parametric decay
lifetimes are possible.  The dark Higgs can decay through gauge kinetic
mixing, with $\epsilon^{-4}$ scaling, or faster if it mixes with 
Standard Model Higgs bosons (e.g. through a
quartic term $\epsilon_\lambda |h_D|^2 |h^2|$).  If the Dark Higgs is stabilized by an accidental symmetry so that its decay is mediated by higher-dimension operators as noted in \cite{Katz:2009qq,Morrissey:2009ur}, the lifetimes can be significantly longer.

For suitable parameter choices, all three decay scenarios (kinetic
mixing, higgs mixing, and higher-dimension operators) lead to
lifetimes $10 \cm \lesssim c\tau \lesssim 10^{15}$ cm, so that
beam-dump and solar limits apply.  The kinetic mixing decay is a
simple and illustrative benchmark, because the decay lifetime and
terrestrial production cross-sections both depend on $\epsilon$ and
the particle masses, with no additional free parameters.  After
discussing the model-independent limits, we will consider
this case explicitly in Section \ref{sec:Example}.  

\section{Beam-Dump Experiments}\label{sec:beamDump}

\begin{figure}
\includegraphics[width=0.5\textwidth]{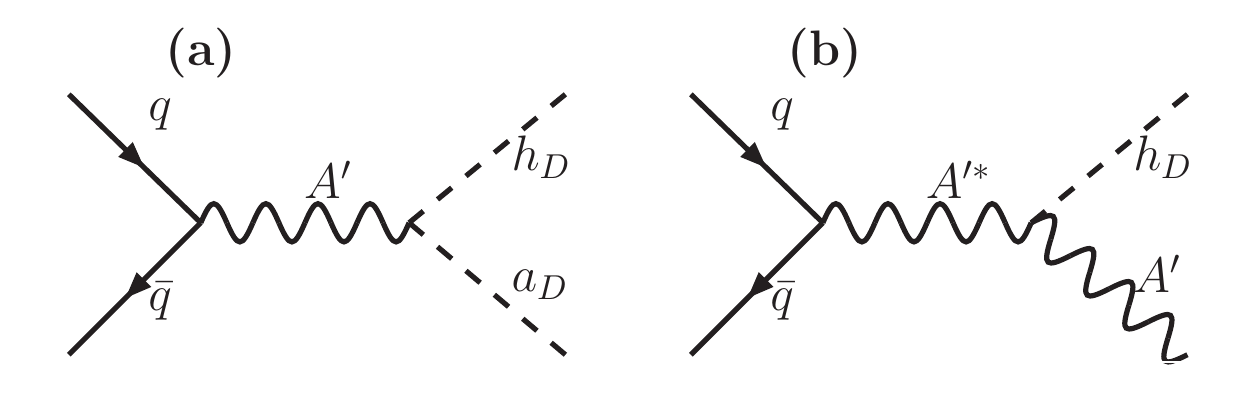}
\caption{\label{fig:dumpFeyn}Feynman diagrams for (a) $A'$ decay into a dark-sector higgs and (b) higgs$'$-strahlung process. }
\end{figure}

Production of dark-sector states in proton beam-dump experiments was
discussed in \cite{Batell:2009di}, which identified three production
modes (meson decays, ``higgs$'$-strahlung'', and resonant $A'$
production). For concreteness, we will assume that a dark sector
scalar higgs $h_D$ is the LLP of interest but the results in this
section are insensitive to details of the model, except for $\epsilon$
characterizing the production cross-section, the $A'$ mass and the
mass and lifetime $c\tau$ of the LLP. We will consider two reactions
for producing scalar dark sector particles: The model-independent
higgs$'$-strahlung process (see Figure \ref{fig:dumpFeyn}(b)) where a
scalar LLP is directly produced, and the significantly larger resonant
process shown in Figure \ref{fig:dumpFeyn}(a), in which the $A'$ can
decay into LLP's.  Data from past experiments such as the axion search
at CHARM \cite{Bergsma:1985qz} have not been interpreted in terms of
these processes anywhere in the literature.

We note that electron beam-dump experiments like the SLAC experiments
E137 \cite{Bjorken:1988as} and E56
\cite{Fryberger:1970tr,Rothenberg:1972yr} are not competitive because
of their lower energy and resulting decreased acceptance~\footnote{The
  sensitivity of E137 to direct decays $A' \rightarrow e^+e^-$ when
  the $A'$ itself is long-lived depends on very forward-peaked
  production kinematics $\theta_{\prod} \sim (m_{A'}/E)^{3/2}$
  \cite{Bjorken:2009mm}, which in this case is erased by the $A'$
  decay.}.  We also note that $h_D$ can be directly produced from pion
decay, or through pion decays into $A's$ followed by decays into
$h_D$. These mechanisms are important in proton beam dumps where pion
production is tremendously large for masses beneath $m_{\pi}$. We
refer the reader to \cite{Batell:2009di} for further analysis.

\subsection{Limits from CHARM Axion Search}

The CHARM axion search \cite{Bergsma:1985qz} had a $4 \times 4~
\rm{m}^2$ detection area, extending over a 35 m decay region, 
2.5 m deep and located 65 m behind a shielded proton target.  
A veto counter was placed in front of the detection volume, so that the search 
was sensitive only to exotic-particle decays within the detector volume.  In all, $N_p=2.4\times
10^{18}$ protons at an energy of 400 GeV were dumped on a copper
target.  No candidate $e^+e^-$, $\mu^+\mu^-$, or $\gamma\gamma$ events
were observed. 

We estimate event rates in the approximation that they are
dominated by collisions in the first nuclear interaction length
$X^{nuc}$ of the material:
\bea
N_X &&\approx (\sigma_n(X) \times A ) \frac{N_0 X^{nuc.}}{A\times 1 \rm{g/mol}} N_p, \nonumber \\
&& = \sigma_n(X) \times 2.0\times 10^{8} \rm{pb}^{-1},
\eea
where $\sigma_n(X)$ is the per-nucleon cross-section for the process
of interest, $A=63.5$ is the average atomic mass of Copper and $N_0$
is Avogadro's constant.

\begin{figure}[htbp]
\includegraphics[width=0.45\textwidth]{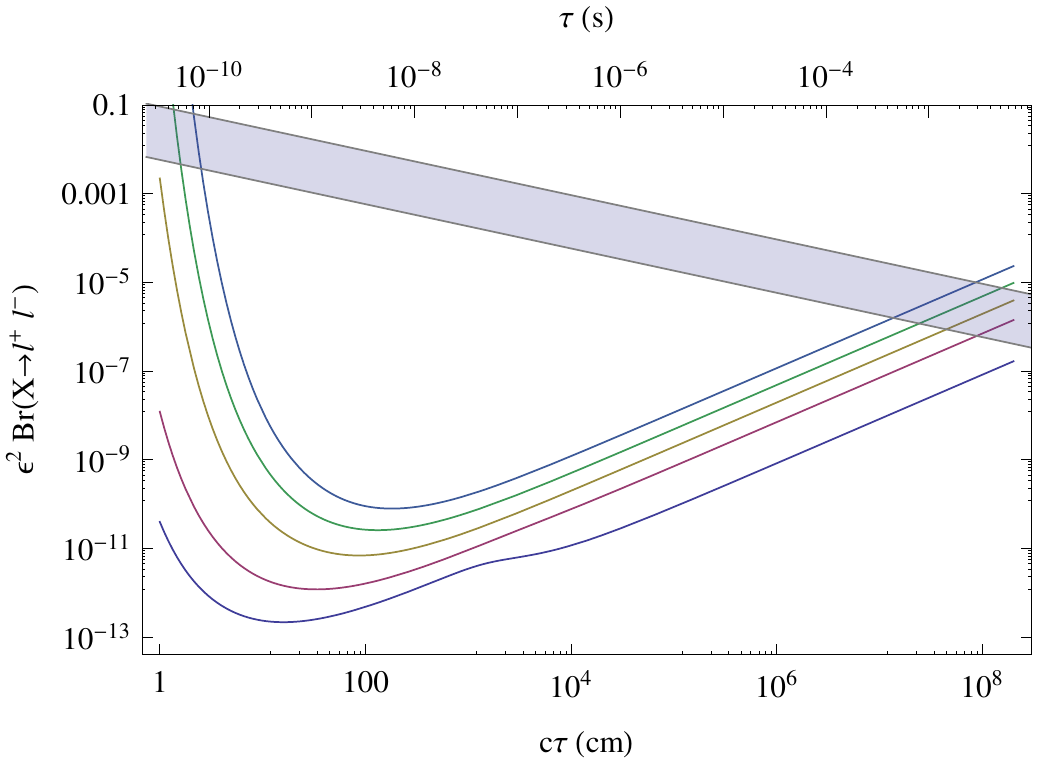}
\caption{Limits on $\epsilon^2 Br(X\rightarrow \ell^+\ell^-)$, as
  a function of decay length assuming resonant production of a vector
  $A'$, with subsequent decay to $X$.  From bottom to top, curves
  correspond to ten events expected (none were observed) for $m_{A'}=$
  0.6, 1, 2, 3, 4 GeV and $m_{h_D} = 0.4 m_{A'}$.  The regions above
  the curves are excluded.  The gray band represents the expected band
  for the scalar model of Section \ref{sec:Example}, with the
  upper and lower lines corresponding to $(m_{A'},m_{h_D})=(0.6,
  0.24)$ GeV and $(4,1.6)$ GeV, respectively.  The width of the band
  comes predominantly from the opening of kaon decay channel at higher
  masses.}
  \label{fig:CHARM_epsLim}
\end{figure}

We have modeled production in MadGraph \cite{Alwall:2007st} and used a
Monte Carlo to determine geometric and lifetime acceptances.  In the
mass range $0.5 \GeV < m_{A'} <4$ GeV, the cross-section for resonant
$A'$ production for a 400 GeV incident proton is well approximated by
\be \sigma(m_{A'}) \sim 20 \pb \left(\f{\epsilon}{10^{-3}}\right)^2
\left( \f{m}{1 \GeV}\right)^{-3} \ee
(for masses above 4 GeV, $\sigma(m_{A'}) \sim <\sigma(4 \GeV) \times
(m_{A'}/4\GeV)^6$).  In addition, the angular acceptance varies
considerably with the $A'$ mass.  The typical energies of $A'$
produced resonantly were $\approx 40-50$ GeV (with 10--20\% exceeding
100 GeV); this is sufficient for a significant fraction of decay
products to travel in the direction of the detector.  Taking into
account angular acceptance and decay probability,
Fig. \ref{fig:CHARM_epsLim} shows the resulting limits on $\epsilon^2
Br(X\rightarrow \ell^+\ell^-)$ for $\ell=e,\mu$ as a function of LLP
lifetime for various $A'$ masses.  These limits were obtained assuming
a decay $A' \rightarrow h_D a_D$ with $m_{h_D}=m_{a_D} = 0.4 m_{A'}$ ,
but are not very sensitive to the detailed decay mode away from mass
thresholds.

\begin{figure}[htbp]
\includegraphics[width=0.45\textwidth]{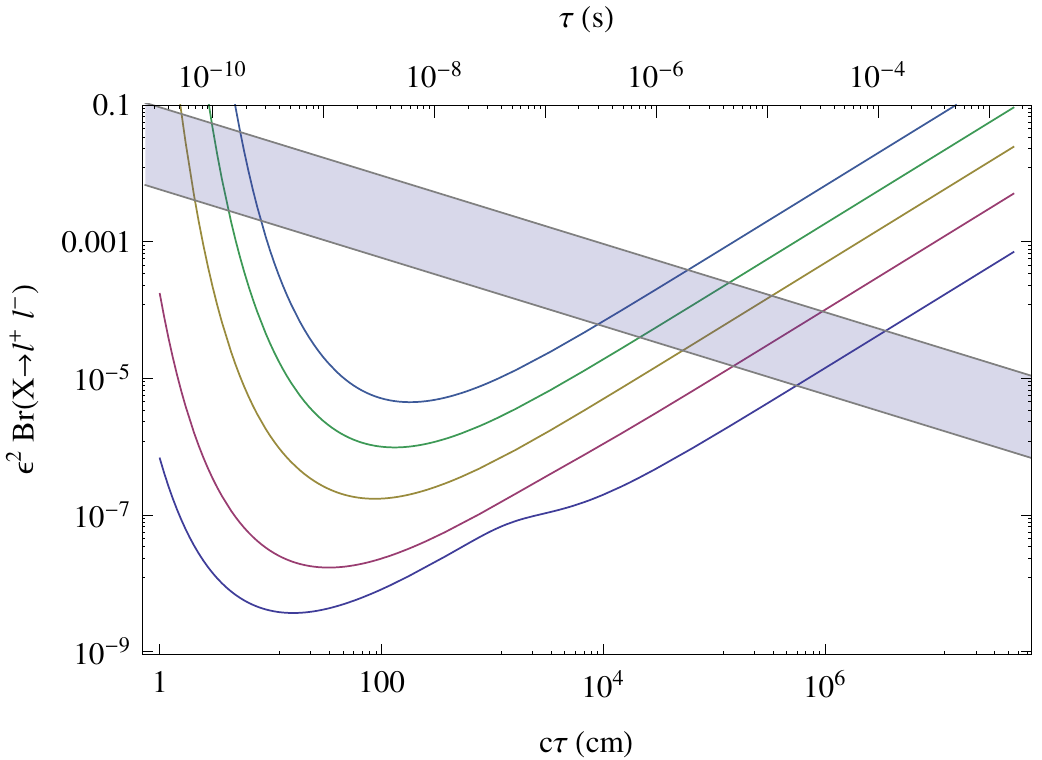}
\caption{ Limits on $\epsilon^2 Br(X\rightarrow \ell^+\ell^-)$ as
  a function of decay length, assuming radiation of $X$ off an
  off-shell $A'$ as in the higgs$'$-strahlung process of
  \cite{Batell:2009yf}.  From bottom to top, curves correspond to ten
  events expected (none were observed) for $m_{A'}=$ 0.6, 1, 2, 3, 4
  GeV and $m_{h_D} = 0.4 m_{A'}$.  The regions above the curves are
  excluded.  The gray band represents the expected band for the scalar
  model of Section \ref{sec:orientation}, with the upper and lower
  lines corresponding to $(m_{A'},m_{h_D})=(0.6, 0.24)$ GeV and
  $(4,1.6)$ GeV respectively.  The width of the band comes
  predominantly from the opening of kaon decay channel at higher
  masses.}
  \label{fig:CHARM_epsLimB}
\end{figure}

For the higgs$'$-strahlung reaction, the cross-section is smaller by 4--5 orders of magnitude due to the  combined effects of a weak coupling, phase space, and decreasing parton luminosity.  For our benchmark $m_{h_D}=0.4 m_{A'}$ and $0.5 < m_{A'} < 10$ GeV, the cross-section can be parametrized as  
\be
\sigma(h_DA') \approx (0.001 \pb ) (m/\GeV)^{-3} \left( 1+(m/4 \GeV)^7\right)^{-1}
\ee
for $\alpha_D=\alpha_{EM}$. Limits on $\epsilon^2 Br(X\rightarrow
  \ell^+\ell^-)$ for this reaction are shown in Figure \ref{fig:CHARM_epsLimB}. 

\section{Dark Matter Annihilation in the Sun and Earth}\label{sec:solar}

We now move on to consider DM capture and annihilations in the Sun~\cite{Press:1985ug,Silk:1985ax} and the Earth~\cite{Freese:1985qw,Krauss:1985aaa}, and the corresponding limits on the production rate and lifetimes of LLPs. We begin by reviewing the capture process of DM in celestial objects and the expected annihilation rates in relation to model parameters. Direct annihilation into SM particles can only be detected by looking for high energy neutrinos coming from the Sun and such searches have been discussed in the literature extensively. Annihilations into long-lived neutral particles, however, may have very different signatures which we now consider. First, they may again be observed in neutrino detectors if the LLPs decay into muons for example. Second, they may be observed in electron detectors such as FERMI. Third, gamma-ray observations, such as EGRET, Milagro and FERMI can detect the radiation associated with electronic decays of the LLPs or their direct decay into photons. This wide range of possible probes of LLP decays is particularly welcome since the precise nature of these particles, if they exist, is still largely unconstrained.  

\subsection{Production Rate from Capture and Annihilation of DM in the Sun and the Earth}\label{sec:solar-production}

We consider several mechanisms for Solar and Earth capture and use the model described by Eq. (\ref{eq:Lag}) as a benchmark.
We first describe the well studied case of elastic scattering of DM against matter and the associated capture rate. We then examine inelastic scattering and the corresponding modified capture rate. In equilibrium, the annihilation rate of DM is directly related to the capture rate. However, the fraction of annihilation into LLPs is model-dependent and so we parametrize it.

The capture rate depends on the scattering cross-section of DM against nuclei. In the models we consider, the scattering is mediated through $A'$ exchange and is given by~\cite{ Pospelov:2007mp},
\be
\sigma_{\chi n} = \frac{16\pi Z^2\alpha \alpha_d \epsilon^2 \mu^2_{ne}}{A^2 m_{A'}^4} 
\label{sigmaChiN}
\ee where $Z$ and $A$ are the atomic and mass number,
respectively. Since the elastic scattering of DM against nuclei is
constrained to be $\lesssim 10^{-43}{\rm cm^2}$ by direct detection
experiments~\cite{Angle:2007uj,Ahmed:2008eu}, the mixing is bounded to
be $\epsilon \lesssim 10^{-6}$ for $m_{A'} \sim \GeV$. These bounds
are dramatically weakened if DM predominantly scatters inelastically
off nuclei as in \cite{TuckerSmith:2001hy}. For example, DM-nucleon
cross-sections $\sigma_{\chi n} \sim 10^{-39}{\rm cm^2}$ can be
compatible with direct detection limits and the DAMA modulation signal
for splittings $\sim 100$ keV \cite{Chang:2008gd}.

\subsubsection*{Elastic Capture}

Elastic capture in celestial objects proceeds through the basic mechanism discussed in \cite{Griest:1986yu,Gould:1987ir}. The scattering can be spin-independent and spin-dependent. Both types of processes are constrained by direct-detection experiments, but the latter much less so. 

If spin-independent scattering cross-sections saturate bounds from direct-detection experiments~\cite{Angle:2007uj,Ahmed:2008eu}, the capture rates are 
\begin{eqnarray}
\label{eqn:CSI_sun}
C^{\rm SI}_{\odot} &=& 1.4 \times10^{21} \mbox{ s}^{-1}\left(\frac{\TeV}{\mX} \right)^{2/3},\\
\label{eqn:CSI_earth}
C^{\rm SI}_{\oplus} &=& 4.1 \times10^{11} \mbox{ s}^{-1}\left(\frac{\TeV}{\mX} \right)^{2/3},
\end{eqnarray}
in the Sun and Earth, respectively~\cite{Gondolo:2004sc}. The spin-dependent cross-section for proton coupling is most stringently constrained by KIMS~\cite{Kim:2008zzn} and setting the cross-section below the experimental bound we find, 
\begin{equation}
C^{\rm SD}_{\odot} = 4.7\times 10^{25}\mbox{ s}^{-1}\left(\frac{100\GeV}{\mX} \right)^{3/2}.
\end{equation} 

\subsubsection*{Inelastic Capture}

Inelastic DM (iDM) was proposed in \cite{TuckerSmith:2001hy} as an explanation for the discrepancy between the annular modulation signal reported by the DAMA collaboration and the null results reported by other experiments. The basic premise of this scenario is that dark matter scatters dominantly off nuclei into an excited state with splittings, $\sim 100\keV$, similar to the kinetic energy of DM in the halo. With this hypothesis, the DAMA data is nicely fit with WIMP-\textit{nucleon} cross-sections of $\sigma_{n\chi} \gtrsim 10^{-40}{\rm~cm^2}$ (see \cite{Chang:2008gd} for details). 
While interactions with matter in the Earth are kinematically forbidden,  
the Sun's gravity provides the necessary kinetic energy to overcome the inelastic threshold and allow for capture of DM in the Sun.

\begin{figure}[h]
\begin{center}
\includegraphics[scale=.62]{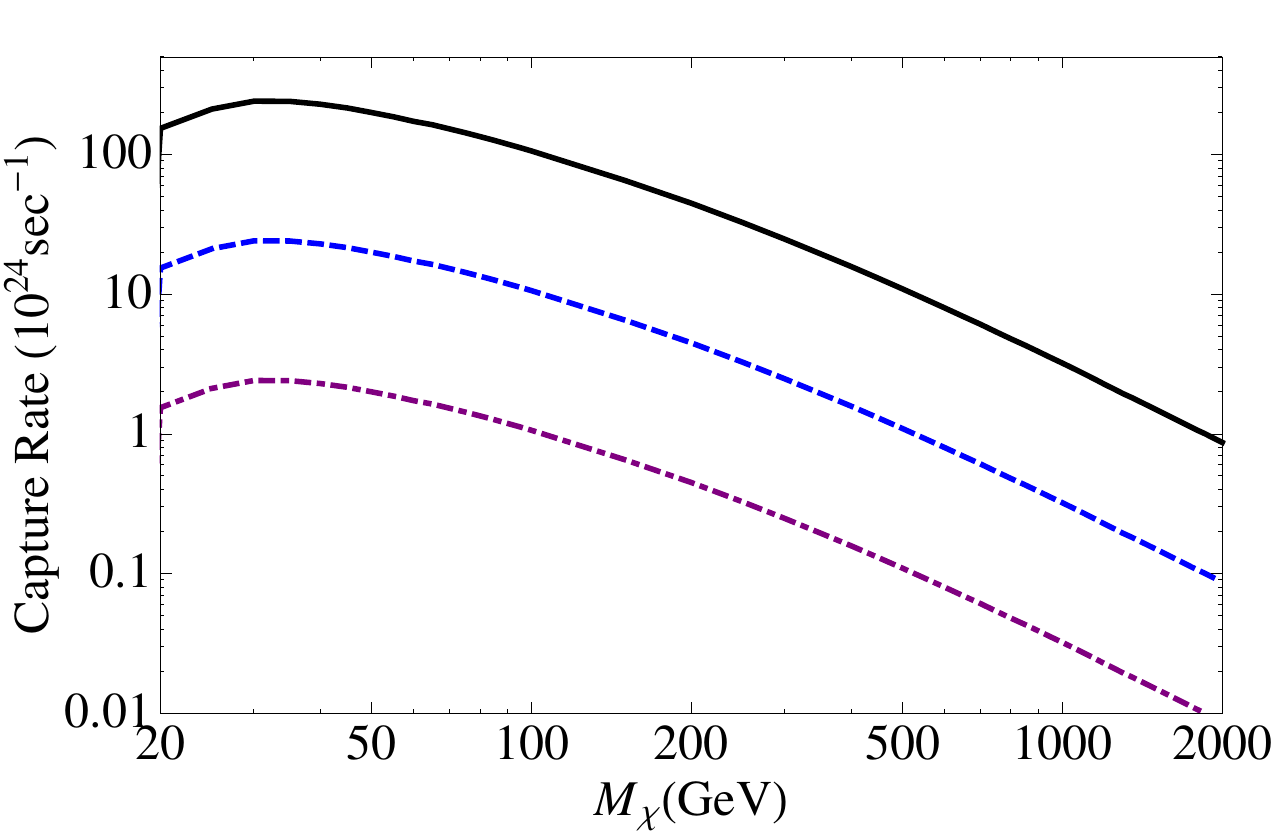}
\end{center}
\caption{The inelastic capture rate of DM in the Sun against the DM mass. The black (solid) curve correspond to an inelastic model with $\exE =125\keV$ and  $\sigma_{n\chi}=10^{-39}~{\rm \cm^2}$, the blue (dashed) curve to  $\sigma_{n\chi}=10^{-40}~{\rm \cm^2}$, and the purple (dash-dot) curve to  $\sigma_{n\chi}=10^{-41}~{\rm \cm^2}$. }
\label{fig:CaptureRate}
\end{figure}

The capture rate in the Sun is shown in Fig.~\ref{fig:CaptureRate} for $\exE = 125\keV$ for several values of the WIMP-nucleon cross-section (see \cite{Nussinov:2009ft} for more details). The parameter space is rather limited because above $\delta\sim 500\keV$ there is no element in the Sun against which DM can scatter inelastically. We note that the high capture rates shown in Fig. \ref{fig:CaptureRate} as compared with Eq.~(\ref{eqn:CSI_sun}) are mostly just a result of the higher WIMP-nucleon cross-section allowed in iDM models. 

\subsubsection*{Annihilation Rate}

In equilibrium the annihilation rate of DM is exactly half its capture rate. More generally it is given by~\cite{Jungman:1995df},
\begin{equation}
\label{eqn:AnnRate}
\Gamma_A = \frac{1}{2} C \tanh^2\left(t/\tau_{\scriptscriptstyle {\rm eq}} \right),
\end{equation}
where $\eqtau$ is the equilibrium time and $t\approx 10^{17}{\rm s}$ is the dynamical time of the system. 
In the elastic scattering case, DM can typically scatter in the Sun on a much shorter timescale than it annihilates. Hence, it thermalizes 
with the rest of the matter in the Sun and concentrates in the inner core as it approaches its equilibrium configuration. 
This is not the case for the Earth where $t/\tau_{\scriptscriptstyle {\rm eq}}\ll 1$ and the annihilation rate is suppressed with respect to the capture rate by 2-3 orders of magnitude. 
If the annihilation cross-section enjoys an enhancement due to the Sommerfeld effect then this suppression can be tamed~\cite{Delaunay:2008pc}.

The situation is quite different for iDM since no more than a few inelastic scatterings can take place before it become energetically impossible to scatter any further. If the elastic cross-section is greater than $\sigma_n \gtrsim10^{-47}~{\rm cm^2}$, there are enough elastic collisions to ensure that 
equilibrium is reached. However, if  $\sigma_n \ll 10^{-47}~{\rm cm^2}$, the collision rate becomes so low that the WIMPs do not thermalize in the Sun. In Ref.~\cite{Nussinov:2009ft} it was shown that equilibrium is nevertheless obtained, and we will therefore equate the annihilation rate with half the capture rate in this case as well. 

\subsection{Limits from the Sun and the Earth}\label{sec:solar-constraints}

\subsubsection*{Bounds from Neutrino Detectors}

If the LLP decays into a $\mu^+\mu^-$ pair anywhere between the Sun's surface and the Earth, these muons can be observed in underground neutrino detectors (Super-K, BAKSAL) as well as the neutrino telescopes (Ice-Cube, Antares) in either of two ways. If the decay happens very close to the detector ($\lesssim {\rm km}$) the muon pair may be observed directly through their Cherenkov radiation as nearby tracks. Alternatively, if the LLP decays into a $\mu^+\mu^-$ pair far from the detector then the muons will yield high-energy neutrinos which can be detected as they convert back into a muon in the rock or ice near the detector. 

\begin{figure}[htbp]
\includegraphics[scale=0.55]{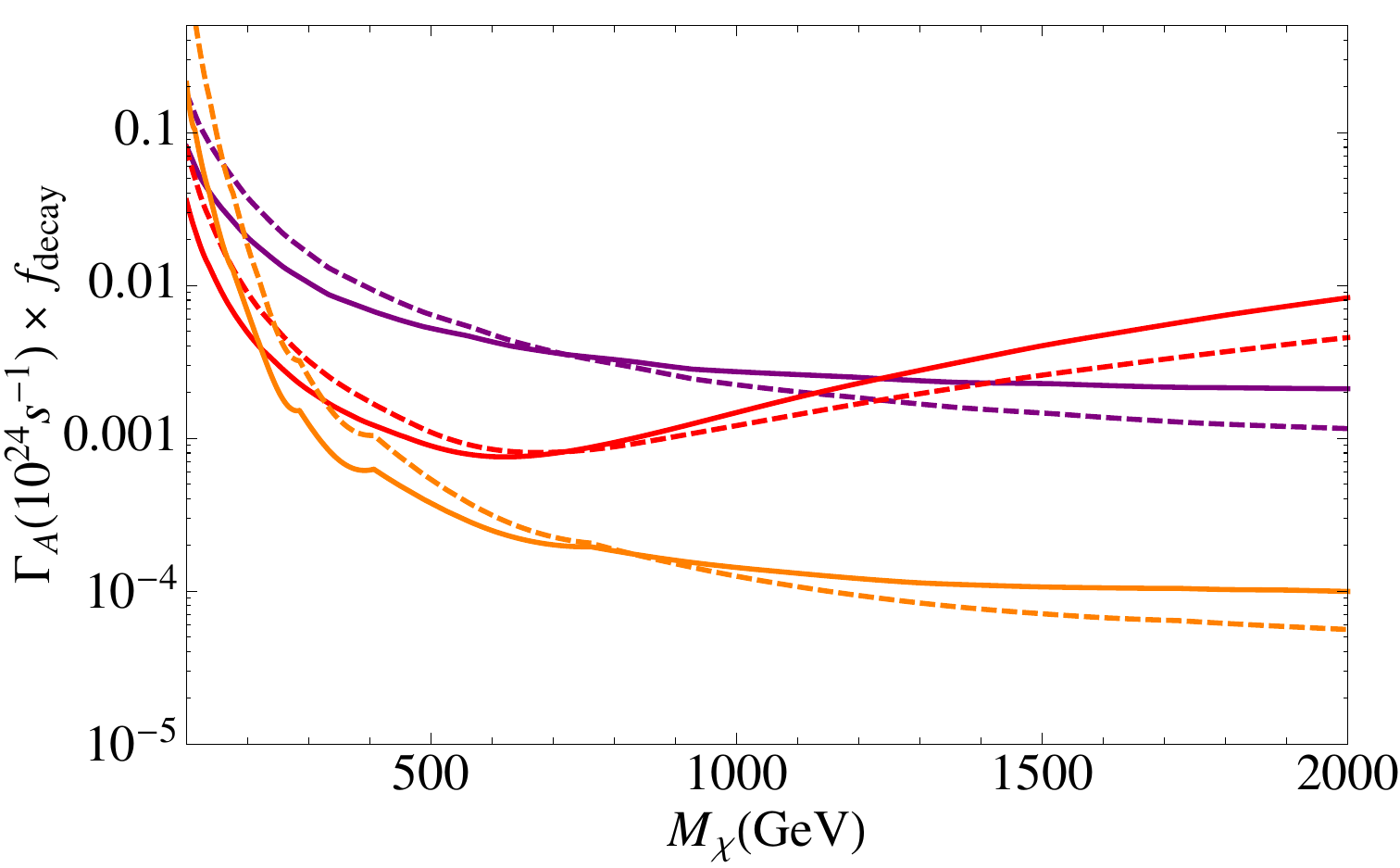}
\caption{\label{fig:sun-LT-R} The purple curves show the current bounds from Super-K on the annihilation rate of DM in the Sun into LLPs assuming 100\% branching ratio into 1-step (solid) or 2-step (dashed) cascades which result in muons. The red (orange) curves are the expected bounds from Antares (Ice-Cube).} 
\end{figure}

The strongest bounds on the flux of upward going muons come
from Super-Kamiokande and are $\mathcal{O}(10^3)~{\rm km^{-2}
  ~yr^{-1}}$~\cite{Desai:2004pq}. Fig. \ref{fig:sun-LT-R} depicts the
bound on the annihilation rate of DM into LLPs that subsequently
decay into muons, and the expected improvement in sensitivity with
Ice-Cube and Antares. Some LLP's scatter or decay inside the Sun,
while others do not decay before reaching Earth, or decay into modes
that do not produce neutrinos.  Thus, the fraction of decays that can be detected near the Earth is,
\begin{eqnarray}
f_{decay}& = &\Br(LLP\rightarrow \mu\rm{'s})  e^{-R_\odot/\ell_{int}} \nonumber \\ 
&& \big( e^{-R_\odot/\gamma c\tau}
 - e^{-{1 AU}/{\gamma c\tau}} \big)
\label{eqn:fDecay}
\end{eqnarray} 
where $\gamma$ is the LLP's boost factor, and 
\begin{eqnarray}
  \ell_{int} & = & (n_\odot \sigma_{int})^{-1}  \approx n_\odot^{-1} \frac{m_{A'}^2}{\pi \alpha \alpha_D \epsilon^2} \nonumber \\
& \simeq & R_{\odot}  \left(\frac{\alpha_D}{\alpha} \right)
\left(\frac{10^{-3}}{\epsilon} \right)^2
\left(\frac{m_{A'}}{100\MeV} \right)^2
\end{eqnarray}

is the LLP mean free path in the Sun.  The suppression
$e^{-R_{\odot}/\ell_{int}} \approx 1$ except for high $\epsilon$ and
  low $m_{A'}$, to which the CHARM analysis is sensitive.  When the
  lifetime becomes much smaller than the Sun's radius, the muons never
  escape and no useful limits can be derived.

\begin{figure}[htbp]
\includegraphics[scale=0.6]{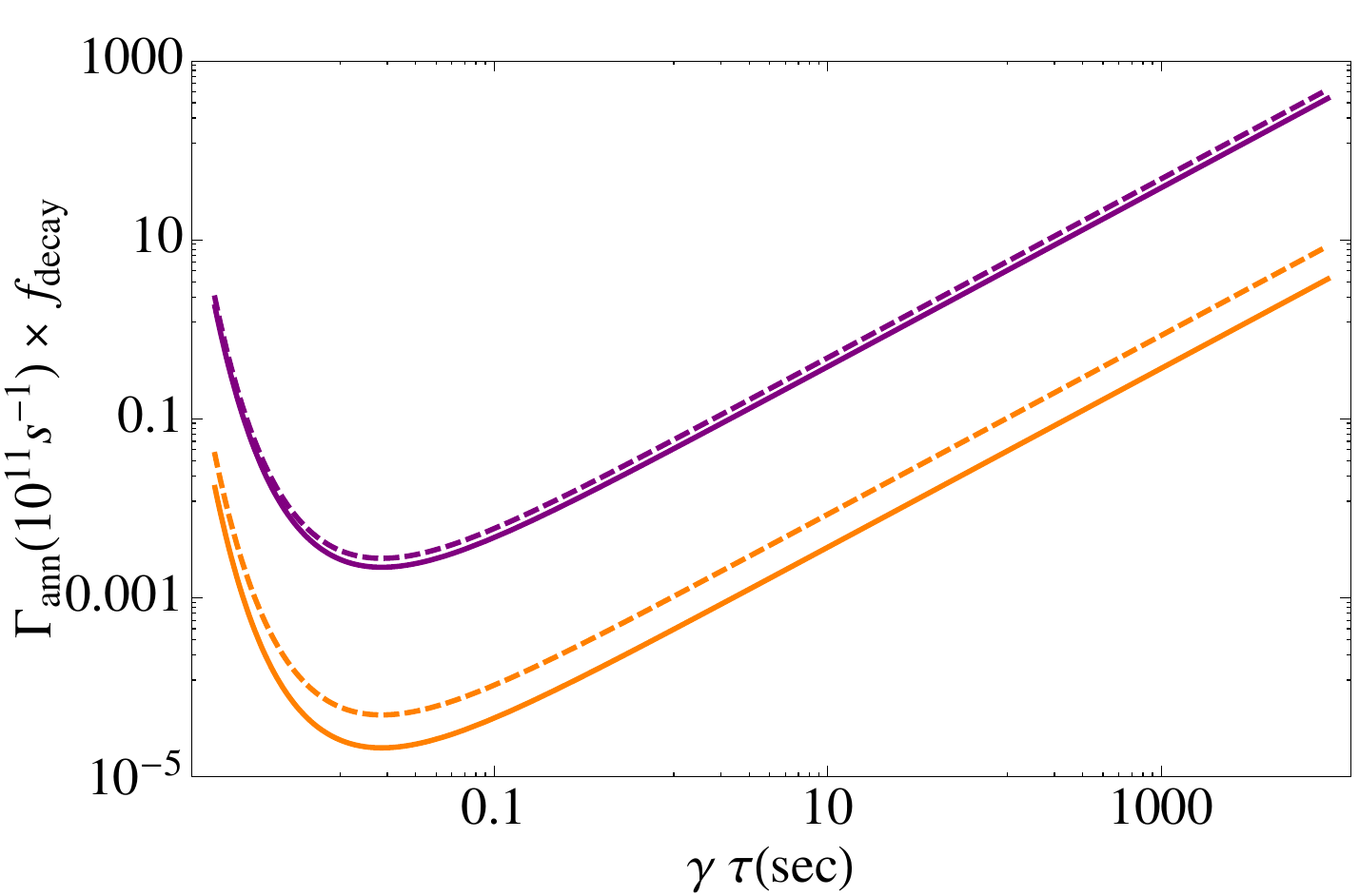}
\caption{\label{fig:earth-LT-R} The purple curves show the constraints from Super-K on the annihilation rate of LLP in the Earth as a function of their lifetime for $\mX = \TeV$ (solid) and $\mX = 0.3\TeV$ (dashed). The orange curves show the expected improvement in sensitivity with Ice-Cube . The mass dependency reflects the improvement for higher masses, but does not take into account possible degradation of the signal due to the colinearity of the muon pair.}
\end{figure}

LLPs produced in the Earth must decay near (<1 km) the detector, or else the muons are stopped in the Earth. In this case, the muons can be observed directly, although one must take into account the special nature of such events. Since the LLPs are very boosted, their decay products (namely the muons) are highly collimated with an opening angle of about $\sim~m_{LLP}/\mX$. The muons are therefore no more than about a meter apart when they reach the detector. We leave it for future work to investigate the efficiency of observing such muon pairs, but there is no obvious reason to suspect that they should be considerably more difficult to see than ordinary events. Hence, the limits from Super-Kamiokande can be used to constrain the annihilation rate of DM in the Earth's core as shown in Fig. \ref{fig:earth-LT-R}. Future results from Ice-Cube may improve on these limits by about two orders of magnitude. We note that such improvements will allow capture rates in the Earth (Eq.~(\ref{eqn:CSI_earth})) to be probed despite the out-of-equilibrium suppression in Eq.~(\ref{eqn:AnnRate}). 

\subsubsection*{Bounds from Electron/Positron Detectors}

Injection of highly energetic leptons between the Sun and Earth can also be observed in dedicated satellite missions that look for electrons/positrons in cosmic rays such as PAMELA~\cite{Adriani:2008zr} and FERMI-LAT~\cite{Abdo:2009zk}. The energy spectrum of such electrons is determined by their production mechanism only and does not suffer from the uncertainties present in galactic propagation models. We consider 1-step and 2-step processes, where the LLP's decay directly into electron/positron pairs or do so via an additional intermediate state, respectively. To obtain the energy spectrum for each type of cascade we assumed the different mass scales involved in the cascade are well separated and that all particles decay isotropically. The relevant formulae can be found in the appendix of \cite{Mardon:2009rc}. Decay of the LLP into a pair of muons is very similar and is in fact well approximated as a 2-step process. 

For such short cascades, the electron energy spectrum is much harder than the cosmic-ray background. Therefore, the strongest constraints are derived from the highest energy bins available. It is then straightforward to derive the bounds on the annihilation rate in the Sun by demanding that the resulting flux is lower than the highest energy bins in FERMI/LAT,  
\begin{eqnarray}
\Gamma_{A}\times f_{decay} &\lesssim& 1.4 \times 10^{20}{\rm ~s^{-1}} \left(\frac{1000\GeV}{\mX} \right)^2\\
\Gamma_{A}\times f_{decay} &\lesssim& 5.1 \times 10^{20}{\rm ~s^{-1}} \left(\frac{1000\GeV}{\mX} \right)^2
\end{eqnarray}
for the 1-step and 2-step cascades, respectively. We note that these are about an order of magnitude stronger than the neutrino constraints from Super-K. 

\subsubsection*{$\gamma$ Ray Constraints}

\begin{figure}[htbp]
\includegraphics[width=0.5\textwidth]{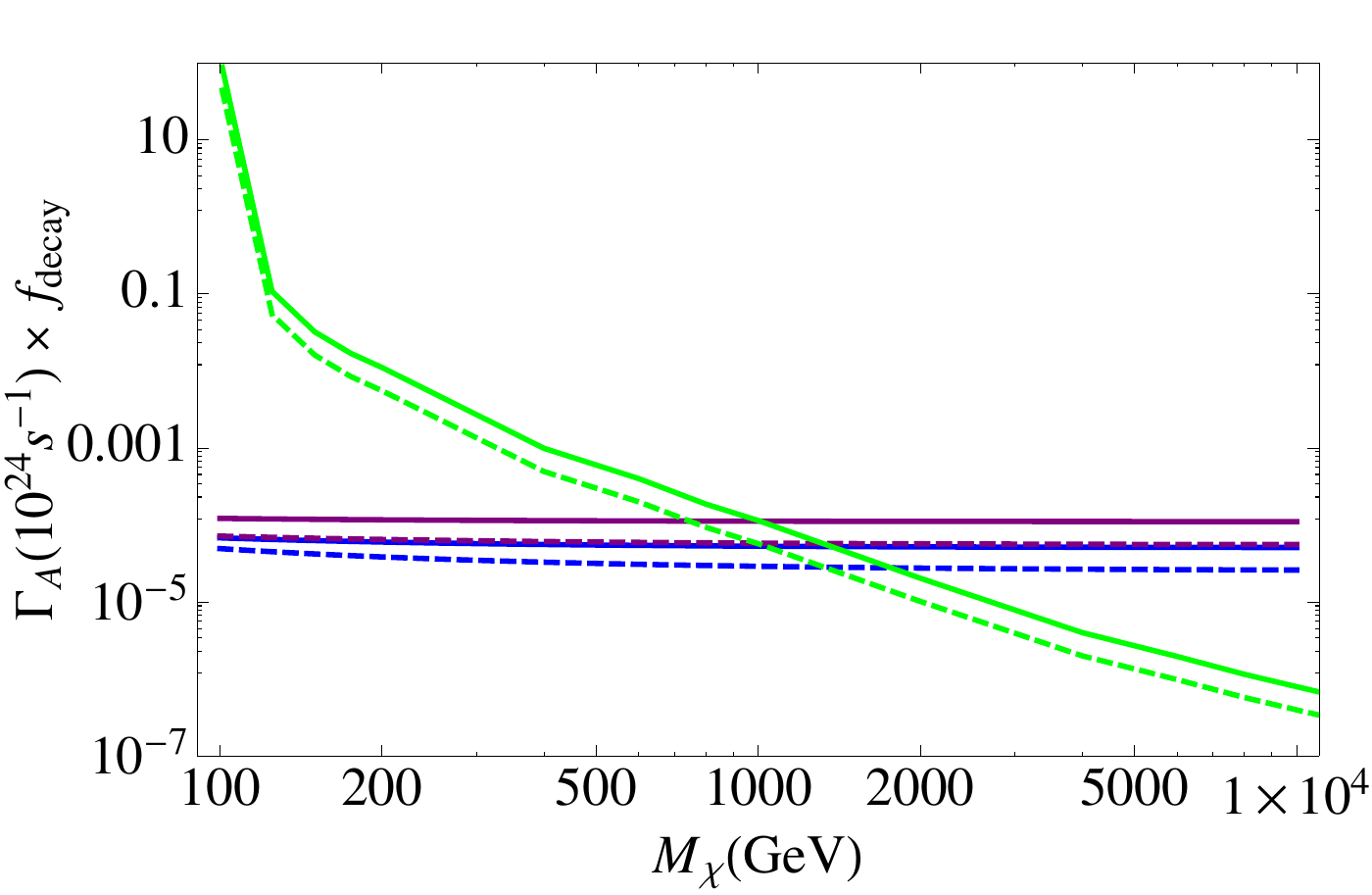}
\caption{\label{fig:GammaRayBounds} FSR constraints on $\Gamma_{\rm ann} \times f_{decay}$ from Milagro (green), EGRET (purple) and Fermi solar data (blue) for one-step (solid) and two-step (dashed) cascade decays, with decay efficiency $f_{decay}$  defined in \eqref{eqn:fDecay}. The bounds were derived as explained in the text.}
\end{figure}

Electronic decays of the LLP will usually lead to final state radiation (FSR) of photons unless the LLP is right above the electron pair threshold. This radiation can be detected by observations of the $\gamma$-ray spectrum of the Sun. Both EGRET and FERMI have looked at the Sun's emission spectrum for $E_\gamma \gtrsim 100\MeV$ and these agree well with the predicted spectrum from inverse Compton scattering of the Sun's light against cosmic rays~\cite{Orlando:2008uk, Giglietto}. We can therefore derive bounds on the electronic decays of LLPs by demanding that the resulting FSR is smaller than the observed gamma ray spectrum. As in the eletronic flux, the FSR energy spectrum is harder than that observed and so the highest energy measurements ($E_\gamma \gtrsim 1\GeV$) actually provide the most stringent limits. For that purpose, the measurement of the very high energy photon spectrum made by Milagro are also useful~\cite{Atkins:2004qr}. To obtain a bound on the flux we require the expected number of events in Milagro to be less than 4791. The resulting bounds on the annihilation rate from the different observations are shown in Fig. \ref{fig:GammaRayBounds}. 

\section{Example: Dark-Sector Scalars}\label{sec:Example}

In this Section we discuss the implications of the previously
identified limits for a particular class of long-lived particles:
Higgs-like dark scalars $h_D$ that decay through loop
diagrams with a rate controlled by the gauge kinetic mixing
$\epsilon$.  We then discuss the dependence of these contours on
variation of model assumptions.
 
In a single-Higgs model the width of the light scalar
$h_D$ to each lepton species is, 
\be
\Gamma_{h_D \rightarrow \ell^+ \ell^-} = \frac{2\alpha_D
    \alpha^2 \epsilon^4}{\pi^2} \frac{I_2(m_\ell, m_{A'},
  m_{h_D})^2}{4} \frac{m_\ell^2}{m_V^2} m_h,
\label{higgsWidth}
\ee
 where $I_2$ is a loop integral defined in \cite{Batell:2009yf}, and
 very nearly equal to $2 \beta^3$, where $\beta = (1-4
 m_\ell^2/m_{h_D}^2)^{1/2}$.  This formula must be corrected by
 mixing angles in a model with two or more Higgses in the dark sector,
 but given our complete ignorance of the dark sector's structure we
 ignore these and take \eqref{higgsWidth} as a benchmark.  For decays
 above the muon threshold, this gives
\bea
c\tau \approx & (2\times 10^7 \cm) \times \frac{1}{Br(h_D \rightarrow \mu^+\mu^-)}
\frac{\alpha}{\alpha_D} \nonumber \\
& \times \left(\frac{10^{-3}}{\epsilon}\right)^4
\left( \frac{m_{A'}}{1 \GeV}\right)^2 \left(\frac{1 \GeV}{m_{h_D}}\right).
\eea
We take the ratio of hadronic to muonic widths from a
spectator model, as in \cite{Gunion:1989we} (we use the parameter
choices of  \cite{McKeen:2008gd} with $r=1$) --
the $\mu^+\mu^-$ branching fraction varies between 20 and 50\%  below
the muon threshold, and falls to 1--2\% above the kaon threshold.

\begin{figure}[htbp]
\includegraphics[width=0.48\textwidth]{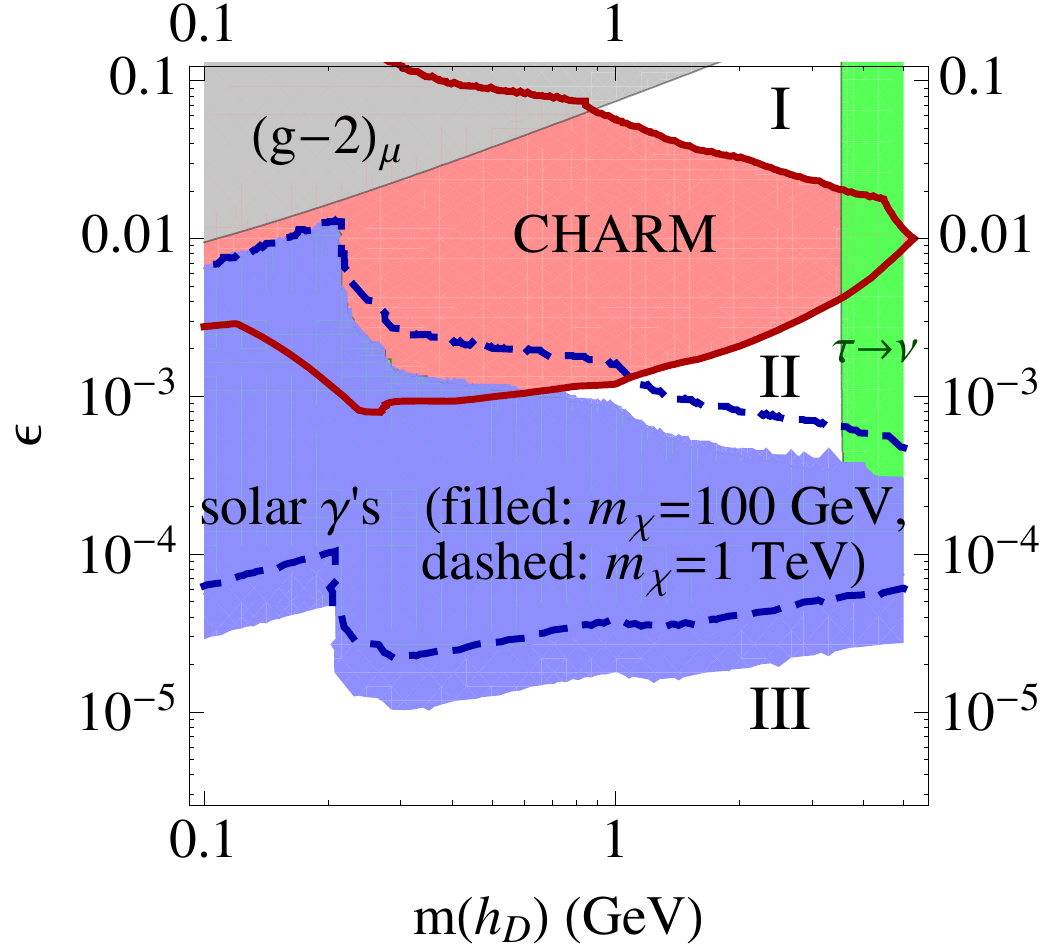}
\caption{\label{fig:higgsConstraints} Combined limits on a higgs-like
  scalar $h_D$ produced in $A'$ decays, with leptonic width
  \eqref{higgsWidth}, $m_{A'} = 2.5 m_{h_D}$, and $\alpha_D = \alpha$.
  The gray region in the upper left is the model-independent limit on
  $\epsilon$ for given $A'$ mass from $(g-2)_\mu$ (see
  \cite{Pospelov:2008zw}).  The red region corresponds to the limit
  from direct $A'$ production in CHARM, with $h_D$ decays into
  electrons or muons (see Sec \ref{sec:beamDump}).  Where this limit
  is obscured by others, the outline of the CHARM exclusion is drawn
  in dark red.  The blue and green regions are solar limits (Sec.~
  \ref{sec:solar-constraints}) relevant if weak-scale dark matter
  couples to the $A'$, and kinetic mixing mediates its scattering off
  ordinary matter.  The blue and green regions are excluded by photon
  and neutrino flux observations, respectively, for a benchmark dark
  matter mass $m_\chi = 100$ GeV.  The dashed blue line outlines the
  analogous exclusion from photon flux when $m_{\chi}= 1$ TeV.  The
  neutrino limits are controlled by the $\tau$ threshold and are
  insensitive to the dark matter mass.  Possible strategies for
  exploring regions I, II, and III are discussed in Section
  \ref{sec:discussion}.}
\end{figure}

The terrestrial direct production limits
and those from DM annihilation in the Sun (assuming annihilation
products include the $A'$ and/or $h_D$) are summarized in Figure
\ref{fig:higgsConstraints}, where for definiteness we have assumed
$m_{h_D}=m_{a_D}=0.4 m_{A'}$, and $\alpha_D = \alpha$.

The most constraining terrestrial experiment is the CHARM beam dump,
which searched for $e^+e^-$ and $\mu^+\mu^-$ events and found no
candidates in $2.4\times 10^{18}$ protons dumped.  These results imply
that $\epsilon \gtrsim 10^{-3}$ are excluded for gauge boson masses up
to a few GeV.  Larger $\epsilon \sim 0.1-1$ is constrained by other
measurements, specifically the muon $g-2$.  

If weak-scale dark matter ($m=$100 GeV--1 TeV) annihilates through the
$A'$, smaller $\epsilon$ can be probed by measurements of the solar
electron, photon, and neutrino flux.  For $m_{h_D} < 2 m_\tau$, the
photon flux from final state radiation of dark Higgses that escape the
Sun is most constraining, with sensitivity to annihilation rates $\sim
10^{19}-10^{20}/\rm{s}$, four to five orders of magnitude below the
expected capture rate for $\sigma_{\chi n}=10^{-40} \cm^2$.  We assume
the capture rate is determined by the scattering cross-section
\eqref{sigmaChiN} from exchange of an $A'$, with no mixing-angle
suppression.  In this case, the solar production rate, like the
terrestrial beam-dump production rate, scales as $\epsilon^2$, leading
to a very similar shape of exclusion region (blue region for
$m_{DM}=100$ GeV, dark blue dashed line for $m_{DM}=1$ TeV).

Unlike muons, pions, and kaons, the $\tau$ decays promptly, producing
neutrinos that can escape the Sun even if the dark-sector Higgs decays
promptly.  Thus dark matter scenarios above this mass threshold are
constrained (green vertical band in Fig. \ref{fig:higgsConstraints}), with
little sensitivity to dark matter mass.

As we have noted, the decay lifetimes of Higgs-like scalars are quite
model-dependent, and can be significantly longer or shorter than those
quoted.  This would qualitatively affect both beam-dump (red) and
solar FSR (blue) constraints.  The upper limits on $\epsilon$
correspond to typical $\gamma c\tau$ comparable to the thickness of
the beam-dump shielding (or $R_{\odot}$), with only logarithmic
sensitivity to production rate.  These shift up (down) for longer
(shorter) lifetimes.  The lower limits are determined equally by the
lifetime (which typically exceeds the size of the experiment) and
production rate, so it is somewhat less affected by variation of the
decay length.

The solar constraints also depend on various assumptions about dark
matter.  The dependence on dark matter mass, shown in the
Fig. \ref{fig:higgsConstraints} (dashed blue line), arises mostly from
the resulting boost of the late-decaying Higgses.  In models with a
reduced DM-matter scattering cross-section, the luminosity of the Sun
in dark matter annihilation is reduced and the resulting FSR limits
are weakened.  We note that a mixing angle weakens the lower-$\epsilon$ limit significantly, but has only logarithmic impact on the
high-$\epsilon$ sensitivity limit.



\section{Discussion}\label{sec:discussion}
Dark sectors coupled to the Standard Model through gauge kinetic
mixing may contain an array of macroscopically long-lived particles
(LLP's) in addition to the kinetically mixed gauge boson.  Gauge
bosons can escape detection in accelerator experiments if they decay
into LLP's with lifetimes between 10 cm and $10^{15}$ cm, rather than
to Standard Model fermions.  LLPs with lifetimes in this range would
escape collider detectors unobserved, but are sufficiently short-lived
that BBN constraints are largely irrelevant.

Remarkably, two classes of data constrain these scenarios across these
fourteen decades in lifetime.  The proton beam dump experiment CHARM places
strong constraints on scalar LLP's with $10 \lesssim c\tau \lesssim
10^8$ cm decaying into muons or electrons. If dark matter annihilates
into dark sector LLP's, then Solar capture of dark matter results in
limits on LLP's with $10^7 \cm \lesssim c\tau \lesssim 10^{15}$ cm for
similar decay modes.

These limits each depend on different assumptions and production
mechanisms, so they cannot be combined into a robust
exclusion. Nonetheless, the case of a Higgs-like boson in the dark
sector decaying through kinetic mixing allows an illustrative
comparison.  The sensitivity of solar limits --- when applicable ---
nearly matches on to the parameter region excluded by CHARM.
Together, these constraints disfavor $A'$ decays to scalars,
pseudoscalars and other long-lived particles in the parameter region
best suited to explain the DAMA modulation signal through iDM
\cite{Katz:2009qq,Alves:2009nf,Morrissey:2009ur}.  In particular, if
LLP's decay through higher-dimension operators, lowering the mass
scale of these operators to evade BBN constraints may bring them into
conflict with experimental data.

Given these limits and the remaining unconstrained parameter regions, our
analysis suggests several promising searches with existing data and
small-scale experiments:
\begin{description}
\item[Prompt Vector Decays to $\ell^+\ell^-$:] Remarkably, for
  $\epsilon \sim 10^{-5}-10^{-3}$, $A'$ decays into LLPs are
  \emph{much more} constrained than decays directly into Standard
  Model states.  This observation compounds the motivation for a
  sharply defined search program to constrain direct $A'$ decays to
  $\ell^+\ell^-$ \cite{Bjorken:2009mm,Reece:2009un}.
\item[Searches in $e^+e^-$ Collider Data:] LLP lifetimes too short to be
  observed in a beam-dump experiment ($\gamma c\tau \lesssim 10$m at
  $E_{LLP}\sim 50 \GeV$, e.g. region I of Fig.~\ref{fig:higgsConstraints}) give
  rise to observable displaced decays within the detectors of  $e^+e^-$
  collider experiments such as Babar, Belle, KLOE, and BES-III.
  Moreover, such lifetimes arise from $\epsilon \gtrsim 10^{-3}$ for which many
  thousands of events are possible in $\gamma + A'$ and $h_D A'$
  production.  The most visible such decay may be into $\mu^+\mu^-$
  (or $\pi^+\pi^-$) pairs reconstructing a displaced vertex, which may
  be accompanied by a photon, invisible particles, and/or additional
  $e$, $\mu$, or $\pi$ pairs.
\item[High-Energy and High Intensity Beam Dumps:] High-energy beam
  dumps with cumulative charges on target exceeding a coulomb may
  extend sensitivity into region II of
  Fig. \ref{fig:higgsConstraints}.  Neutrino factories with near
  detectors such as MINOS have geometries well-suited for beam-dump
  LLP searches \cite{Batell:2009di,Essig:2009aa}, and integrated
  luminosity $\sim 10^3$ larger than that of CHARM.
\item[Solar Gamma Ray and Electron Measurements:] LLP lifetimes $\sim
  10^{3}-10^{5}$ seconds (region III in
  Fig. \ref{fig:higgsConstraints}) are too short to disrupt light
  element abundances, but only a fraction of LLP's can decay between
  the Sun and the Earth, so very sensitive solar measurements are
  needed to probe these scenarios.  Solar gamma-ray observations above
  10 GeV from ACTs and FERMI can explore this region, as can
  searches for electronic signals correlated with the Sun
  \cite{Schuster:2009aa}.
\item[Milli-Charged Particle Searches:] Electron beam-dump experiments
  searching for milli-charged particles (e.g.
  \cite{Prinz:1998ua,Jaros:1995kz}) may be sensitive to LLP
  scattering off a nuclear target $N$ in the active detector volume,
  such as $h_D+N\rightarrow A'+N$.  Searches in existing data may be
  sensitive to $\epsilon \gtrsim 10^{-3} (100 \MeV/m_{A'})$ and
  $m_{LLP} \lesssim 100$ MeV.
\end{description}


{\it Note added: While this work was brought to completion two related works} \cite{Batell:2009zp}, {\it and } \cite{Meade:2009mu}{\it appeared.}
\acknowledgments

We thank Nicola Giglietto and Simona Murgia for useful discussions of
FERMI bounds from the Sun, Shmuel Nussinov for discussions regarding
Milagro, Steven Sekula for discussions about BaBar search limits.  We
also thank Brian Batell,  Patrick Meade, Michele Papucci, Maxim Pospelov, Adam Ritz , Tomer Volansky, and especially Neal Weiner for useful discussions.  We thank Rouven Essig for valuable comments on the draft.  PS and NT thank NYU and the IAS for hospitality during the
completion of this work. Likewise, IY thanks SLAC and the KIAS for
their hospitality.  PS is supported by the US DOE under contract
number DE-AC02-76SF00515.  IY is supported by the James Arthur
fellowship.


\bibliographystyle{apsrev}
\bibliography{LongLivedConstraints}
\end{document}